\documentclass[twocolumn,showpacs,preprintnumbers,amsmath,amssymb]{revtex4}
\usepackage{graphicx}
\usepackage{dcolumn}
\usepackage{bm}

\begin{document}

\title{Temporal and spatial interference in molecular above-threshold
ionization with elliptically polarized fields}

\author{XuanYang Lai$^{1,2}$ and C. Figueira de Morisson Faria$^{1}$}
\affiliation{$^{1}$Department of Physics and Astronomy, University
College London, Gower Street, London WC1E 6BT, UK\\$^{2}$State Key
Laboratory of Magnetic Resonance and Atomic and Molecular Physics,
Wuhan Institute of Physics and Mathematics, Chinese Academy of
Sciences, Wuhan 430071, China}
\date{\today}

\begin{abstract}
We investigate direct above-threshold ionization in diatomic
molecules, with particular emphasis on how quantum interference is
altered by a driving field of non-vanishing ellipticity. This
interference may be either temporal, i.e., related to ionization
events occurring at different times, or spatial, i.e., related to
the electron emission at different centers in the molecule.
Employing the strong-field approximation and saddle-point methods,
we find that, in general, for non-vanishing ellipticity, there will
be a blurring of the temporal and spatial interference patterns. The
former blurring is caused by the electron velocity component
perpendicular to the major polarization axis, while spatial
interference is washed out as a consequence either of $s-p$ mixing,
or of the temporal dependence of the ionization prefactor. Both
types of interference are analyzed in detail in terms of electron
trajectories, and specific conditions for which sharp fringes occur
are provided.
\end{abstract}

\pacs{33.80.Rv, 33.80.Wz, 42.50.Hz} \maketitle

\section{Introduction}

Atoms and molecules interacting with a strong laser field may be
ionized by absorbing more photons than necessary. This highly
nonlinear phenomenon is called above-threshold ionization (ATI) and
has attracted increasing attention since its 1979 discovery by
Agostini and co-workers \cite{Agostini1979PRL}. Recently, impressive
progress has been achieved in the study of ATI. For example, great
emphasis has been placed on employing ATI as a tool for the
attosecond imaging of molecular orbitals
\cite{Meckel2008Science,Kang2010PRL}. Imaging applications are based
on the physical mechanism behind this phenomenon, in which the
active electron (i) is released in the continuum, (ii) is
accelerated by the field, and (iii) reaches the detector without
returning or after rescattering with its parent molecule. The former
and the latter scenario are known as direct ATI, and as rescattered
or high-order ATI (HATI), respectively. Quantum mechanically, this
electron may be released along several pathways, and the
corresponding transition amplitudes will interfere.

Previous studies of ATI have shown that for the strong-field
ionization of molecules, there are two kinds of quantum interference
displayed in the ATI spectrum. In the first scenario, for a given
energy, the electronic wave packet may reach the continuum at events
occurring at different times in one cycle of laser field
\cite{Becker2002AdvAtMolOptPhys}. Hence, the transition amplitudes
associated with ionization along different orbits in the time domain
will interfere. This leads to a fringe structure in the ATI spectrum
\cite{Milo2012PRA,Bauer2005PRA}, which is present regardless of
whether the target is an atom or a molecule.  For that reason, we
will refer to this type of interference as ``temporal interference".
Recently, ATI spectra with unusual two-photon separation in the
direction perpendicular to the laser polarization
\cite{Korneev2012PRL} have been revealed due to the temporal
interference.

In the second scenario, due to the multi-core structure of
molecules, there may be electron emission from spatially separated
centers, which results in a multi-slit like interference pattern in
the ATI spectrum. In the context of linearly polarized fields and
diatomic molecules, spatial-interference effects have been widely
studied
\cite{Fano1966PR,Walter1999JPB,Lei2002PRA,selsto2005PRL,Baltenkov2012JPB,bohm2000PRL,Busuladzic2008PRA,Li2012PRA,Carla2011PRA}.
Such studies range from simplified molecular models to exact
three-dimensional time-dependent calculations. Employing the
molecular strong-field approximation (SFA), a generalized spatial
interference condition for diatomic molecules that accounts for the
orbital geometry and $s-p$ mixing has been derived
\cite{Busuladzic2008PRA,Li2012PRA,Carla2011PRA}. For such fields,
however, the dynamics are mainly confined along the field
polarization axis.

Elliptically polarized laser fields exhibit a nonvanishing
electric-field component perpendicular to the major polarization
axis, which affects the motion of the electron in the laser field.
Hence, they have been proposed as a resource for controlling
strong-field phenomena by varying the ellipticity. For example, it
is known that elliptical polarization has great significance as a
tool to produce isolated attosecond pulses \cite{ottawa06} and to
increase the contribution of long orbits in the HATI process
\cite{Lai2013PRL}. So far, the ellipticity dependence of ATI has
been widely studied since the mid 1990s. Most studies, however,
focus on the influence of the field ellipticity on the angular
dependence of the ionization rate of ATI
\cite{Eckle2008NatureP,Busuladzic2009PRA,Goreslavski2004PRL,Staudte2009PRL},
while the imprints left on the ATI spectrum by a nonvanishing
driving-field ellipticity have received little attention.

In an early paper \cite{Paulus1998PRL}, features related to the
above-mentioned temporal interference have been identified
experimentally for direct ATI.  Therein, it was found that the
photoelectron yield as a function of the driving-field ellipticity
behaves in two very distinct ways for HATI and direct ATI energy
ranges. While in the former case this yield decreases monotonically
for larger driving field ellipticity, following the predictions of
classical models, for direct ATI the electron ellipticity
distributions display several ``shoulder" structures. These
structures have been related to the quantum temporal interference
between the two sets of orbits that contribute to ATI in this energy
region. However, how the ellipticity influences the two orbits still
remains unexplored to a great extent.

In this work, we are interested in how the driving-field ellipticity
influences the above-mentioned two kinds of interferences in the ATI
spectra of diatomic molecules, exemplified by N$_{2}$ and Ar$_{2}$.
Throughout, we focus on direct ATI, for which temporal interference
effects are also prominent. Employing the molecular SFA and
saddle-point methods, we find that, in general, the temporal
interference in the ATI spectra will become blurred in the
elliptically polarized field. For electron momenta along the major
polarization axis, however, the temporal interference remains clear
and the positions of the interference minima shift towards lower
photoelectron energies with increasing ellipticity. These features
are well explained in terms of the ellipticity dependence of the
initial electron velocity along each orbit, and we show that the
downshift of the interference minima is responsible for the
``shoulder" structure observed in the ellipticity distributions of
ATI electrons in Ref.~\cite{Paulus1998PRL}.

Furthermore, for the spatial interference, clear interference minima
which do not depend on the ellipticity are found for N$_2$ if only
$s$ or $p$ states are taken. However, if $s-p$ mixing is introduced,
sharp fringes are only present for linear polarization if the
molecule is aligned perpendicular to the field axis. For all other
cases, these minima are blurred even for linear polarization. For
Ar$_2$, the ATI spectra display several clear minima caused by the
spatial interference. The reasons for the differences between
N$_{2}$ and Ar$_{2}$ molecules are discussed in detail in terms of
the shapes of their highest occupied molecular orbitals (HOMO). In
this view, we propose a necessary condition to observe the spatial
interference in the ATI spectrum in experiments.

This paper is organized as follows. In Sec.~II, we provide a brief
discussion of the model, including how to obtain the ATI amplitude
and two-center interference conditions. Subsequently, we present the
ATI spectra of the N$_2$ and Ar$_2$ molecules in the elliptically
polarized laser field, and discuss how quantum interference is
altered by the driving field of non-vanishing ellipticity. Finally,
in Sec. IV our conclusions are given. Atomic units (a.u.) are used
throughout unless otherwise indicated.
\section{Model}

\label{model}

\subsection{Transition Amplitude}

Within the molecular SFA, the transition amplitude for direct ATI
from the initial bound state $\left\vert \psi _{0}\right\rangle $ to
a
final Volkov state with drift momentum $\mathbf{p}$ is given by \cite%
{Carla2002PRA,Becker2002AdvAtMolOptPhys,KFR}
\begin{equation}\label{Mpdir}
M_{d}(\mathbf{p})=-i\int_{-\infty }^{\infty }dt\left\langle
\mathbf{p}+\mathbf{A}(t)\right. |-\mathbf{r}\cdot \mathbf{E}%
(t)|\left. \psi _{0}\right\rangle \exp [iS_{d}(\mathbf{p},t)],
\end{equation}
where
\begin{equation}
S_{d}(\mathbf{p},t)=-\frac{1}{2}\int_{t}^{\infty }[\mathbf{p}+%
\mathbf{A}(\tau )]^{2}d\tau +I_{p}t  \label{action}
\end{equation}
is the semiclassical action. In Eqs.~(\ref{Mpdir}) and
(\ref{action}), $I_{p}$ gives the ionization potential, and
$\mathbf{E}(t)$ and $\mathbf{A}(t)$ denote the laser electric field
and the vector potential, respectively. For simplicity, we assume
that all the molecular structure is contained in the ionization
prefactor
\begin{equation}
d_{ion}=\left\langle \mathbf{p}+\mathbf{A}(t)\right. |-\mathbf{r}%
\cdot \mathbf{E}(t)|\left. \psi _{0}\right\rangle .
\label{prefactor}
\end{equation}%
This is the most widespread assumption when performing the SFA
modeling of strong-field phenomena in molecular systems
(\cite{Carla2010PRA,Milo2006PRA,bohm2000PRL,Madsen2004JPB,Chu2005PRA};
see also the reviews \cite{Lein_Review,Carla_Review}), and implies
that the active electron leaves from the geometrical center of the
molecule. In this work, we employ the single-active orbital
approximation, i.e., we assume that the initial state $\left\vert
\psi _{0}\right\rangle $ is the HOMO. This orbital is written as the
linear combination of atomic orbitals (LCAO), and the core is
assumed to be frozen. This latter approximation is reasonable for
molecules with heavy nuclei \cite{Madsen2006}. Explicitly,
\begin{equation} \label{molecule}
\psi _{0}(\mathbf{r})=\sum_{\alpha }c_{\alpha }\left[ \psi _{\alpha
}(\mathbf{r}+\frac{\mathbf{R}}{2})+(-1)^{l_{\alpha }-m_{\alpha
}+\lambda _{\alpha }}\psi _{\alpha
}(\mathbf{r}-\frac{\mathbf{R}}{2})\right] ,
\end{equation}%
where the wavefunctions $\psi _{\alpha }(\mathbf{r})$ give the
atomic orbitals, $\mathbf{R}$ is
the internuclear distance, $l_{\alpha }$ is the orbital quantum number, $%
m_{\alpha }$ is the magnetic quantum number and $\lambda _{\alpha
}=m_{\alpha }$ for gerade (g) and $\lambda _{\alpha }=m+1$ for
ungerade (u) orbital symmetry \cite{Wahl1964JCP}.

\subsection{Saddle-point equation and temporal interference}

For sufficiently high intensity and low frequency of the laser
field, the temporal integrations in the amplitude (\ref{Mpdir}) can
be mathematically evaluated by the saddle-point method with high
accuracy \cite{Carla2002PRA,Becker2002AdvAtMolOptPhys}, i.e., we
seek solutions such that the action (\ref{action}) is stationary.
For the specific case of an elliptically polarized field confined to
the $xy$ plane, the saddle-point equation to be solved reads as
\begin{equation}
\frac{\lbrack p_{x}+A_{x}(t)]^{2}}{2}+\frac{[p_{y}+A_{y}(t)]^{2}}{2}+I_{p}=0,
\label{saddle}
\end{equation}%
where $p_{x}$ and $p_{y}$ denote the final momentum in the
$\hat{\mathbf{x}}$ direction and in the $\hat{\mathbf{y}}$
direction, respectively. Physically, Eq.~(\ref{saddle}) ensures the
conservation of energy at the ionization time
$t$. In terms of the solutions $t_{s}$ of Eq.~(\ref%
{saddle}), the transition amplitude  (\ref{Mpdir}) can be written as
\begin{equation}
M_{d}(\mathbf{p})\propto \sum_{s}\mathcal{C}_s(t_s)d_{ion}\exp
[iS_{d}(\mathbf{p},t_{s})],  \label{sfa_saddle}
\end{equation}%
where the prefactors
\begin{equation}
\mathcal{C}_s(t_s)=\sqrt{\frac{2\pi i}{\partial
^{2}S_{d}(\mathbf{p},t_{s})/\partial t_{s}^{2}}}
\end{equation}
are expected to vary much more slowly than the action at each saddle
for the above-stated approximation to hold \cite{Bleistein}. For
each final momentum $\mathbf{p}$, there are two solutions of
Eq.~(\ref {saddle}) within one period of the laser field, whose
contributions will interfere. This interference is present
regardless of whether the target has one or more centers, i.e., for
atoms and molecules, and has been extensively studied in the
literature for linearly polarized driving fields
\cite{Korneev2012PRL,Milo2012PRA,Bauer2005PRA}.

In this work, we employ the monochromatic elliptically polarized
laser field
\begin{equation}
\mathbf{E}(t)=\frac{\omega A_{0}}{\sqrt{1+\xi
^{2}}}(\hat{\mathbf{x}}\sin \omega t-\xi \hat{\mathbf{y}}\cos \omega
t),  \label{Et}
\end{equation}%
with vector potential
\begin{equation}
\mathbf{A}(t)=\frac{A_{0}}{\sqrt{1+\xi ^{2}}}(\hat{\mathbf{x}}\cos
\omega t+\xi \hat{\mathbf{y}}\sin \omega t),  \label{At}
\end{equation}%
where $\xi $ is the laser ellipticity. In this case, all cycles are
equivalent and the amplitude $M_{d}(\mathbf{p})$ can be intuitively
understood as the coherent superposition of the contributions of the
two orbits, which may add constructively or destructively. \ For
vanishing electron momenta, these orbits will start at two
consecutive maxima of the driving field $\mathbf{E}(t)$.\ These
times will approach each other as the momentum increases.
Throughout, we will refer to the orbits starting in the first and
second half cycle of the field (\ref{Et}) as orbit 1 and orbit 2,
respectively. For a discussion of these solutions for linearly
polarized monochromatic fields see \cite{KopoldPhD,SF2010,SFS2012}
\footnote{The latter two references discuss the
recollision-excitation with subsequent ionization process in
nonsequential double ionization. In this case, the saddle-point
equation obtained for the second electron is formally identical to
that encountered in direct ATI.}.

\subsection{Spatial interference condition}

Due to the two-center structure of the diatomic molecule, one may
expect electron emission from spatially separated centers in the
molecule. This results in minima and maxima in the ATI spectra,
which, in the present model, are implicit in the ionization dipole
matrix element $d_{ion}$. Using
the length gauge with field-dressed molecular bound states \cite%
{DressedStates}, the ionization prefactor (\ref{prefactor}) can be
rewritten as
\begin{equation}\label{prefactor2}
\begin{split}
d_{ion}=  -\sum_{\alpha }c_{\alpha } & [ \exp (i\mathbf{p}\cdot \mathbf{R}%
/2)+(-1)^{l_{\alpha }-m_{\alpha }+\lambda _{\alpha }} \\
& \exp (-i\mathbf{p}\cdot \mathbf{R}/2)] \mathbf{E}(t)\cdot
i\partial _{\tilde{\mathbf{p}}}\psi _{\alpha }(\tilde{\mathbf{p}})
\end{split}
\end{equation}
where $\psi _{\alpha }(\tilde{\mathbf{p}})$ is the Fourier transform
of the atomic orbitals $\psi _{\alpha }(\mathbf{r})$ given by
Eq.~(\ref{molecule}) with
$\tilde{\mathbf{p}}=\mathbf{p}+\mathbf{A}(t)$. The
two-center
interference is related to the terms in the bracket in Eq.~(\ref{prefactor2}%
). This generalized interference condition accounts for $s-p$ mixing
and the orbital geometry, and reduces to simpler interference
conditions if only $s$ or $p$ states are present. For example, 
the terms in bracket reduce to $2\cos
(\mathbf{p}\cdot \mathbf{R}/2)$ for $s$ states and gerade (g) symmetry.
Thus, interference minima are present if
\begin{equation}
\mathbf{p}\cdot \mathbf{R}=(2k+1)\pi ,  \label{sstate}
\end{equation}%
where $k$ denotes an integer number. Similarly, the interference
minima are given by the condition
\begin{equation}
\mathbf{p}\cdot \mathbf{R}=2k\pi  \label{pstate}
\end{equation}
for $p$ states and gerade symmetry. If, on the other hand, the HOMO
has ungerade (u) symmetry, the above-stated scenario is reversed,
i.e., the destructive interference condition for $p$ states is given
by Eq. (\ref{sstate}), while Eq. (\ref{pstate}) holds for $s$ states.

In this paper, the molecular axis is kept in the plane $xy$ spanned by
the driving-field ellipticity. The frame of reference
($\hat{\textbf{x}}'$,$\hat{\textbf{y}}'$,$\hat{\textbf{z}}'$) of the
molecule is rotated by an angle $\eta$  in this plane, defined with
regard to the major polarization axis of the laser field. This means that
$\hat{\textbf{x}}=\sin \eta \hat{\textbf{y}}'+\cos \eta
\hat{\textbf{x}}'$, $\hat{\textbf{y}}=\cos \eta
\hat{\textbf{y}}'-\sin \eta \hat{\textbf{x}}'$, and
$\hat{\textbf{z}}= \hat{\textbf{z}}'$.

The wave functions $\psi _{\alpha }(\mathbf{r})$ in Eq.
(\ref{molecule}) will be approximated by a Gaussian basis set
\begin{equation}
\psi _{\alpha }(\mathbf{r})=\sum_{v}b_{v}(x')^{\beta _{x'}}(y')^{\beta
_{y'}}(z')^{\beta _{z'}}\exp (-\zeta _{v}r^{2}),
\end{equation}%
where the coefficients $c_{\alpha}$, $b_{v}$, $\beta_{x',y',z'}$ and
the exponents $\zeta _{v}$ are extracted from the quantum chemistry
code GAMESS-UK \cite{GamessUK}. Throughout, we employ a 6-31G basis
so that only $s$ and $p$ states are considered. For a Gaussian basis
set, the wave function $\psi_{\alpha}  (\tilde{\mathbf{p}})$ in
momentum space can be written as
\begin{equation}
\psi_{\alpha} (\tilde{\mathbf{p}})=\sum_{v}b_{v}\varphi
_{v}(\tilde{\mathbf{p}}).
\end{equation}
Usually, the explicit formula of $\varphi _{v}(\tilde{\mathbf{p}})$
is complicated. However, if only $s$ and $p$ states are used in the
orbital construction, $\varphi _{v}(\tilde{\mathbf{p}})$ is
simplified as
\begin{equation}\label{p_gauess}
\varphi_{v}
(\tilde{\mathbf{p}})=(-i\tilde{p}_{\chi})^{l_{\alpha}}\frac{\pi^{3/2}}{
2^{l_{\alpha}}\zeta_v^{3/2+l_{\alpha}}}
\exp[-\tilde{p}^{2}/(4\zeta_v)],
\end{equation}
where $l_{\alpha}=0,1$ and the subscript $\chi$ indicates the
orientation of the $p$ states employed to construct the orbital. For
instance, a $\sigma$ orbital would require $p$ states along the
$\hat{\textbf{x}}'$ axis, i.e., $\chi=x'$, whereas a $\pi$ orbital
would require $p$ states along the $\hat{\textbf{y}}'$ or
$\hat{\textbf{z}}'$ axis, i.e., $\chi=y'$ or $z'$.

In this work, we will study the N$_{2}$ and Ar$_{2}$ molecules, both
of which have $\sigma$ HOMOs. In combination with Eqs. (\ref{p_gauess}) and
(\ref{saddle}), the prefactor in Eq. (\ref{prefactor2}) can be
specified as
\begin{equation}\label{prefactor3}
\begin{split}
d_{ion} &= i \cos(\frac{\mathbf{p}\cdot
\textbf{R}}{2})\sum_{\alpha(s),v }c_{\alpha }b_{v}
[\tilde{p}_{x}E_{x}(t)+\tilde{p}_{y}E_{y}(t)] \mathcal{D}_{s}+
\\
&  i \sin(\frac{\mathbf{p}\cdot \textbf{R}}{2})\sum_{\alpha(p),v
}c_{\alpha } b_{v} \{   [(\tilde{p}_{x}^{2}-2\zeta_v) \cos\eta
+\tilde{p}_{x}\tilde{p}_{y}\sin \eta]\times
\\&
E_{x}(t) +[(\tilde{p}_{y}^{2}-2\zeta_v) \sin\eta
+\tilde{p}_{x}\tilde{p}_{y}\cos \eta] E_{y}(t)   \} \mathcal{D}_{p}
\end{split}
\end{equation}
where $\mathcal{D}_{s}=
e^{\frac{I_{p}}{2\zeta_v}}\pi^{3/2}/\zeta_v^{5/2}$ and
$\mathcal{D}_{p}=
e^{\frac{I_p}{2\zeta_v}}\pi^{3/2}/(2\zeta_v^{7/2}$). Note that, in
Eq.~(\ref{prefactor3}), the momentum components relate to the
electric field frame of reference.

\section{Results and discussion}

In the results that follow, we will analyze how the above-mentioned
temporal and spatial interference are influenced by a non-vanishing
driving field ellipticity. This analysis will be performed in terms
of electron orbits. For the temporal interference, we will focus on
the N$_{2}$ molecule, which has been widely studied in the
literature, both theoretically and experimentally. For all
parameters employed in Sec.~\ref{temporal}, the structure caused by
electron emission at separate centers gets washed out, so that
temporal interference may be clearly assessed. In our studies of
spatial interference effects (Sec.~\ref{spatial}), we consider the
N$_{2}$ and Ar$_{2}$ molecules. Such molecules have  very different
HOMO geometries and equilibrium internuclear distances.

\subsection{Temporal interference}\label{temporal}

\begin{figure}[tbp]
\begin{center}
\includegraphics[scale=0.3]{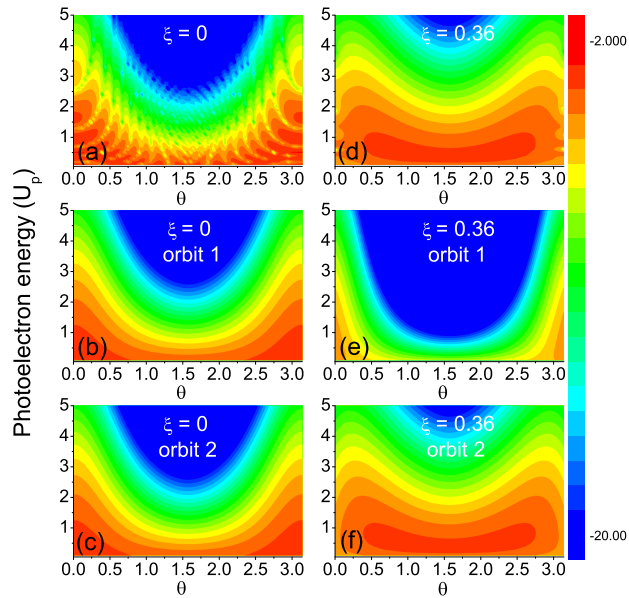}
\end{center}
\caption{(Color online) ATI spectra of the N$_{2}$ molecule, coded
in false colors in the ($E_{\mathbf{p}}$, $\protect\theta $) plane
and plotted in a logarithmic scale. In panels (a) and (d), both
orbits 1 and 2 were included, for $\protect\xi=0$ and
$\protect\xi=0.36$, respectively, while in panels (b), (c), (e) and
(f) the contributions of individual orbits were taken. In panels (b)
and (c), the field has been chosen to be linearly polarized
($\protect\xi=0$), while in panels (e) and (f) the ellipticity is
$\protect\xi=0.36$. The major axis of the polarization ellipse is
along the molecular axis, i.e., for alignment angle $\eta=0$. The
laser intensity and wavelength are $I=2\times 10^{14}$ W/cm$^{2}$
and $\lambda=$800 nm, respectively.} \label{fig1}
\end{figure}

\begin{figure}[tbp]
\begin{center}
\includegraphics[scale=0.3]{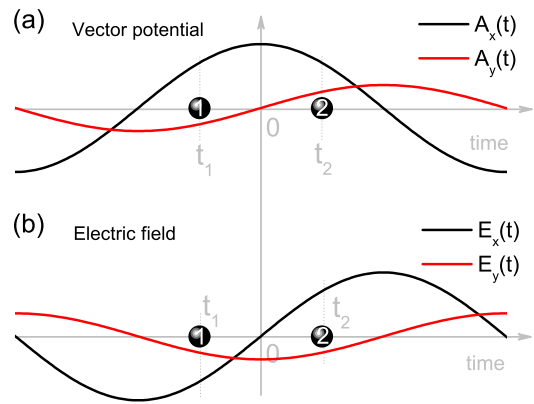}
\end{center}
\caption{(Color online) Time evolution of the two components of the
vector potential $\mathbf{A}(t)=\frac{A_{0}}{\sqrt{1+\xi
^{2}}}(\hat{\mathbf{x}}\cos \omega t+\xi \hat{\mathbf{y}}\sin \omega
t)$ and the electric field $\mathbf{E}(t)=\frac{\omega
A_{0}}{\sqrt{1+\xi ^{2}}}(\hat{\mathbf{x}}\sin \omega t-\xi
\hat{\mathbf{y}}\cos \omega t)$, with $\xi=0.36$ [panels(a) and (b),
respectively]. The two points $t_{1}$ and $t_{2}$ denote the times
associated to the two ionization events $1$ and $2$ that occur
within a field cycle, for an electron with the same final momentum
$\mathbf{p}$. These events are indicated by the colored circles in
the figure. For simplicity, we have defined the time related to a
field crossing $E_x(t)=0$ for the electric-field component along the
major polarization axis as $t=0$. This time also gives a maximum of
the larger vector potential component $A_x(t)$. Note that
$A_x(t_1)=A_x(t_2)$ as both events are equidistant in time from
$t=0$.} \label{vector}
\end{figure}

Figure \ref{fig1} shows the ATI spectra of the N$_{2}$ molecule,
coded in false colors in the ($E_{\mathbf{p}}$, $\theta $) plane,
where $\theta =\angle (\mathbf{R},\mathbf{p})$ denotes the angle
between the internuclear distance and the drift momentum, and
$E_{\mathbf{p}}=\mathbf{p}^{2}/2$ is the kinetic photoelectron
energy upon reaching the detector, in units of the ponderomotive
potential $U_{p}$. We assume that the major axis of the polarization
ellipse is along the molecular axis, i.e., that the alignment angle
$\eta=\angle (\mathbf{R},\mathbf{\hat{x}})$ is vanishing. For linear
polarization [see Fig.~\ref{fig1}(a)], many clear minima can be
observed in the spectra. These minima are due to the destructive
interference between orbit 1 and orbit 2. To illustrate this more
clearly, we calculate the ATI transition probability density
$|M_d(\mathbf{p})|^2$ employing only individual orbits, for the same
parameters as in Fig.~\ref{fig1}(a). These results are displayed in
Figs.~\ref{fig1}(b) and (c), and unambiguously show no minima.

With increasing ellipticity, the interference minima become blurred
and, for most angles $\theta$, completely disappear as the
ellipticity reaches $\xi=0.36$, as shown in Fig. \ref{fig1}(d). In
fact, only the minima along and in the vicinity of  the major
polarization axis ($\theta =0$ and $\pi $) remain visible for this
ellipticity. Sharp minima are only present exactly along the axis
\footnote{In \cite{Paulus1998PRL}, it was shown that, for
$\theta=0$, the saddle-point Eq.~(\ref{saddle}) exhibits two
relevant solutions in the complex time plane for $\xi\leq 0.78$,
whose imaginary parts are identical and which merge into a single
solution around $\xi=0.78$. It is only for these larger
ellipticities that the fringes will vanish for this specific
angle.}.

This behavior is due to the fact that, in general, a non-vanishing
driving-field ellipticity affects the initial electron velocity
along orbit 1 and orbit 2 unequally. The initial velocity of the
electron at the ionization time $t$ is $\mathbf{p}+\mathbf{A}(t)$,
and thus the velocity along the minor polarization axis is given by
$v_{y}=p\sin \theta +A_{y}(t)$. In Fig. \ref{vector}(a), we present
the time evolution of the vector potential $\mathbf{A}(t)$ with
ellipticity $\xi =0.36$. The two times $t_{1}$ and $t_{2}$ represent
the ionization times related to orbit 1 and 2, respectively, and are
equidistant from the electric field crossing $E_x(t)=0$ [see
Fig.~\ref{vector}(b)]. The figure clearly shows that the directions
of the vector potential $\mathbf{A}_{y}$ are opposite at the two
ionization times. Thus if the electron leaves along orbit 1, the
absolute value of its initial velocity $v_{y}$ should be larger than
if it leaves along orbit 2.  This holds except for linear
polarization, i.e., if $A_{y}(t)\equiv 0$, or for electron momenta
along the major polarization axis, i.e., for $\theta =0$ or $\pi $.
The
ionization rate at a specific ionization time $t$ is evaluated by $%
w(t,v_{0})=w_0(t)w_1(t,v_0)$ \cite{Delone1991JOSA}, where
\begin{equation}
w_0(t)=\frac{4(2I_{p})^{2}}{E(t)}\exp \left[
-\frac{2}{3E(t)}(2I_{p})^{3/2}\right] , \label{w0}
\end{equation}%
\begin{equation}
w_1(t,v_0)=\frac{(2I_{p})^{2}}{E(t)\pi }\exp \left[ -\frac{v_{0}^{2}}{E(t)}(2I_{p})^{1/2}%
\right] ,  \label{w1}
\end{equation}%
and $v_{0}=\sqrt{v_{x}^{2}+v_{y}^{2}}$ is the initial velocity.
Hence, the larger the initial velocity $v_{0}$ is, the lower the
ionization rate at that ionization time will be. This implies that,
for elliptical polarization, ionization along orbit 2 will become
larger than ionization along orbit 1. Consequently, the destructive
interference between two electron orbits will disappear in the ATI
spectra. Here we assume that the transverse velocity of the electron
is $p_y=p\sin \theta\leq 0$. If $p_y\geq 0$ the above-stated results
are reversed.

This is in agreement with Figs.~\ref{fig1}(e) and (f), in which we
present the individual contributions to the ATI spectra from orbit 1
and orbit 2, respectively, for ellipticity $\xi =0.36$. In general,
the probability density $|M_d(\mathbf{p})|^2$ associated with orbit
1 is much lower than that related to orbit 2. This is in stark
contrast to what is observed for linear polarization, for which the
two probability densities are identical [see Figs.~\ref{fig1}(b) and
(c)]. However, for drift momentum $\textbf{p}$ along the major
polarization axis ($\theta =0$ and $\pi $), the ATI probability
densities are the same, because the two orbits exhibit the same
initial perpendicular velocity $|v_{y}|=|A_{y}(t)|$. Therefore, the
coherent superposition of the corresponding transition amplitudes
will lead to  clear interference minima in the ATI spectra, as
observed in Fig.~\ref{fig1}(d).

\begin{figure}[tbp]
\begin{center}
\includegraphics[scale=0.3]{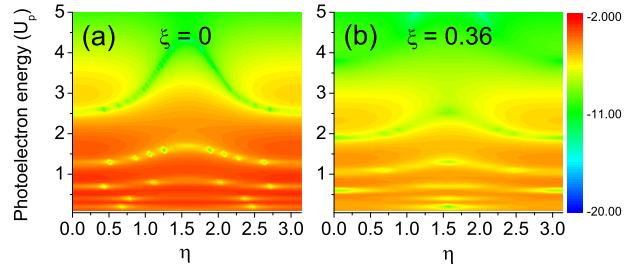}
\end{center}
\caption{(Color online) ATI spectra of the N$_{2}$ molecule along the major
polarization axis  as a function of the alignment angle $\eta$,
coded in false colors in the ($E_{\mathbf{p}}$, $\eta$) plane.
Ellipticity $\xi=0$ for (a) and $\xi=0.36$ for (b). The parameters
of the laser field are the same as in Fig. \protect\ref{fig1}.}
\label{moving}
\end{figure}
\begin{figure}[tbp]
\begin{center}
\includegraphics[scale=0.3]{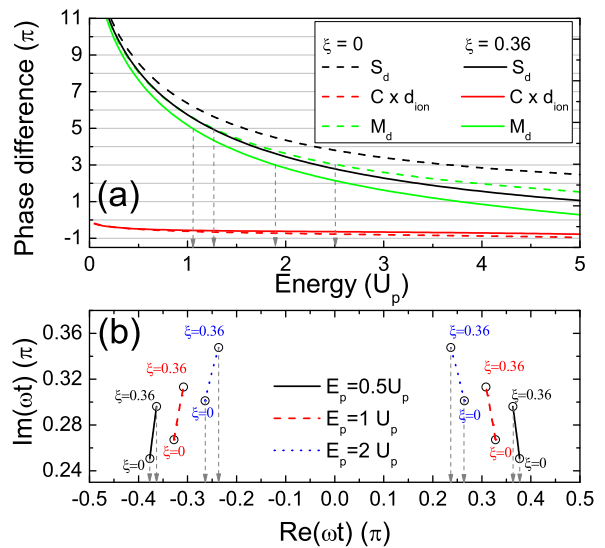}
\end{center}
\caption{(Color online) (a) Phase difference between the
contributions of the two orbits in Eq. (\protect\ref{sfa_saddle})
for parallel alignment $\eta=0$ and momentum along the major
polarization axis ($\theta=0$) with $\protect\xi=0$ (dashed lines)
and $\protect\xi=0.36$ (full lines): action (black), prefactor
(red), and total transition amplitudes (green). The gray arrows
indicate the values of the energy when the total phase difference is
equal to $3\pi$ and $5\pi$ for $\xi=0$ and $\xi=0.36$. (b) Saddle
points associated with the solutions of Eq.~(\ref{saddle}) for the
two orbits as functions of the ellipticity from $\xi=0$ to
$\xi=0.36$, for parallel alignment, momenta along the major
polarization axis, and kinetic energies $E_{p}=0.5U_{p}$ (black full
lines), $E_{p}=U_{p}$ (red dashed lines) and $E_{p}=2U_{p}$ (blue
dotted lines). The gray arrows indicate the real values of the
saddle points for $\xi=0$ and $\xi=0.36$.} \label{Action}
\end{figure}

Next we will further study the interference minima in the ATI
spectra along the major polarization axis. In Fig.~\ref{moving}, we
calculate the ATI spectra of the N$_{2}$ molecule with $\theta=0$ as
a function of the alignment angle $\eta$. The most eye-catching
features in the spectra are the clear minima due to the destructive
interference between the two electron orbits. With increasing
ellipticity, the ATI spectra seem to be squeezed and the positions
of the minima are moving down to lower energies. For example, in
Fig.~\ref{moving}(a) the position of one of the minima is about
$2.5U_{p}$ for $\eta=0$ and $\xi=0$, but it becomes $1.9U_{p}$ in
Fig.~\ref{moving}(b) for $\xi=0.36$. Furthermore, Fig.~\ref{moving}
shows that the downshift of the minima with the field ellipticity is
much more dramatic for higher photoelectron energies than in the
low-energy region.

In order to understand the behavior of the minima in
Fig.~\ref{moving}, we calculate the phase difference between the two
orbits in Eq.~(\ref {sfa_saddle}), which determines the positions of
the interference minima in the ATI spectra.  In
Fig.~\ref{Action}(a), we exhibit the phase difference between the
amplitudes $M^{(1)}_d(\mathbf{p})$ and $M^{(2)}_d(\mathbf{p})$, the
actions $S_{d}(\mathbf{p},t_1)$ and $S_{d}(\mathbf{p},t_2)$, and the
prefactors $\mathcal{C}_1(\mathbf{p}, t_1)d_{ion}^{(1)}$ and
$\mathcal{C}_2(\mathbf{p}, t_2)d_{ion}^{(2)}$ associated with the
two orbits for parallel alignment $\eta=0$, and field polarizations
$\protect\xi=0$ and $\protect\xi =0.36$ (dashed and full lines,
respectively). The figure shows that the phase difference between
$S_{d}(\mathbf{p},t_1)$ and $S_{d}(\mathbf{p},t_2)$ (black lines) is
smaller for $\xi=0.36$ than for $\xi=0$. This is caused by the real
parts of two saddles $t_1$ and $t_2$ approaching each other for
elliptical polarization \footnote{Since the imaginary parts of two
saddles $t_1$ and $t_2$ are the same for a fixed polarization, the
change in the real parts are responsible for the shifts of minima in
the patterns.}, as illustrated in Fig.~\ref{Action}(b). Moreover,
the approaching of the two saddles becomes more significant in the
high-energy than in the low-energy region of the photoelectron
spectra [see the gray arrows in Fig.~\ref{Action}(b)]. Therefore,
the decrease of the phase difference associated with the action is
much more dramatic for higher photoelectron energies than in the
low-energy region. For the prefactor, the phase difference changes
much less markedly with regard to the ellipticity, as shown in the
red lines in Fig.~\ref{Action}(a). This is in agreement with the
assumption of the steepest descent method that the prefactor varies
more slowly with time than the action \cite{Bleistein}.  Thus, the
change in the total phase difference [green lines in
Fig.~\ref{Action}(a)] with the ellipticity is mainly determined by
the action. When the difference in the total phase is equal to
$(2k+1)\pi, k=0,1,2,...$, destructive interference between the two
orbits will occur. In Fig.~\ref{Action}(a), the positions of the
total phase difference with $k=1$ and $k=2$ are marked with arrows.
The two positions are about $2.5U_{p}$ and $1.3U_{p}$ for linear
polarization and move down to about $1.9U_{p}$ and $1.1U_{p}$,
respectively, for elliptical polarization. Moreover, the downshift
of the minima in the high-energy region is more significant than
that in the low-energy region. These results are consistent with the
behaviors of the interference minima observed in Fig.~\ref{moving}.

Furthermore, the behavior of the real parts of the saddle points in
Fig.~\ref{Action}(b) can be explained in terms of the ellipticity
dependence of the initial ionization velocity $v_{0}$. In the
classical limit, the real parts $\mathrm{Re}(t_1)$ and
$\mathrm{Re}(t_2)$ are associated with the classical ionization
times $t_1$ and $t_2$ of an electron in the field (\ref{Et}),
obtained if $I_p \rightarrow 0$. In a linearly polarized laser
field, if an electron is released with drift momentum $\mathbf{p}$
along the major polarization axis, it will reach the continuum with
the vanishing initial velocity $v_{0}=0$. In an elliptically
polarized field, however, the electron should have a nonvanishing
initial velocity in order to compensate the motion induced by the
small component of the field. In the classical limit, the total
initial velocity $\textbf{v}_{0}(t)$ of the electron is
perpendicular to the laser electric field $\mathbf{E}(t)$ at the
corresponding ionization time $t$
\cite{Goreslavski2004PRL,Goreslavski1996ZETF}, i.e.,
$\textbf{E}(t)\cdot\textbf{v}_{0}(t)=0$. The condition
$\textbf{v}_{0}(t)\perp \mathbf{E}(t)$, $t=t_1,t_2$ requires that
$v_x(t_1)>0$  and $v_x(t_2)>0$ for the parameters employed in this
paper. Since
$v_{x}=p+ A_{0} \cos \omega t_{1} /\sqrt{1+\xi^{2}}$ and $A_x(t)>0$
for both $t_1$ and $t_2$ [see Fig.~\ref{vector}(a)], an increase in
the value of $v_{x}$ with the driving-field ellipticity requires
that $t_1$ increases in order to approach the time for which
$A_x(t)$ is maximal, i.e., $t=0$ [see Fig.~\ref{vector}]. Moreover,
for the photoelectrons in the high energy region, the ionization
time is closer to $t=0$ \cite{Becker2002AdvAtMolOptPhys}. Hence, a
larger initial velocity $v_{x}=-E_y/ E_x v_{y}=\xi^2 A_0 \cos \omega
t/\sqrt{1+\xi^{2}}$ will be required to satisfy the condition
$\mathbf{v}_0 \perp \mathbf{E}(t)$. This results in a larger
increase of the ionization time. The same argument applies to orbit
2, with the difference that, because this orbit is located after the
electric-field crossing [see Fig.~\ref{vector}(b)], $t_2$ should
decrease. As a direct consequence, for increasing ellipticity the
two ionization times should approach each other, and  the
approaching of the two times becomes more significant for higher
photoelectron energies than in the low-energy region.  These results
are in good agreement with the two saddle-point solutions as
functions of the driving-field ellipticity shown in
Fig.~\ref{Action}(b) (or see Fig.~4 in Ref.~\cite{Paulus1998PRL}).

It is worth mentioning that, because the interference minima in
Fig.~\ref{moving} move down to lower energy with increasing
ellipticity, for fixed photoelectron energy there will be
fluctuations in the ATI spectrum. This interesting phenomenon has
been observed in experiments \cite{Paulus1998PRL} and was called the
``shoulder" structure in the ellipticity distribution of the ATI
electrons. In our work, the underlying mechanism of the ``shoulder"
structure is explained in terms of the ellipticity dependence of the
initial ionization velocity.

\subsection{Two-center interference}\label{spatial}

\begin{figure}[tbp]
\begin{center}
\includegraphics[scale=0.3]{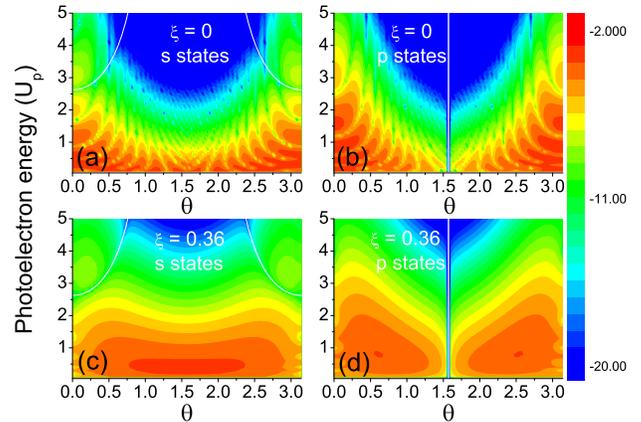}
\end{center}
\caption{(Color online) ATI spectra of the N$_{2}$ molecule computed
using only the s states [panels (a) and (c)] and the p states
[panels(b) and (d)] from its HOMO, coded in false colors in the
($E_{\mathbf{p}}$, $\protect\theta$) plane. The major axis of the
polarization ellipse is along the molecular axis, $\protect\eta=0$.
Ellipticity $\protect\xi=0$ for (a) and (c) and $\protect\xi=0.36$
for (b) and (d). In all panels, the two-center interference minima
are indicated by the white curves in the figure. The remaining field
and molecular parameters are the same as in Fig.~\ref{fig1}.}
\label{ati_sp}
\end{figure}

Finally, we will study the spatial interference due to electron
emission at the two spatially separated centers in N$_{2}$ and
Ar$_{2}$. The former molecule has a $3\sigma _{g}$ HOMO with a high
degree of $s-p$ mixing. For the latter molecule, the HOMO is a
$5\sigma _{u}$ orbital, in which the $p$ states dominate.
Furthermore, Ar$_{2}$ exhibits a very large equilibrium internuclear
distance ($R=7.2$ a.u.), so that, according to
Eq.~(\ref{prefactor2}), several two center minima are expected to be
present in the momentum range of interest.

In Fig.~\ref{ati_sp}, we display the ATI spectra of  N$_{2}$, if
only the contributions from the $s$ or $p$ states to the HOMO are
taken. These spectra are coded in false colors in the
($E_{\mathbf{p}}$, $\theta $) plane. The major axis of the
polarization ellipse is chosen to be parallel to the molecular axis,
i.e., at the alignment angle $\eta=0$. Fig.~\ref{ati_sp}(a) shows
that for the $s$-state contributions and linear polarization, apart
from the minima caused by the temporal interference, there are other
two minima, whose positions satisfy the two-center interference
condition, Eq.~(\ref{sstate}), shown by the white lines in the
figure.  With increasing ellipticity [see Fig.~\ref{ati_sp}(c)], the
minima from the temporal interference are blurred as discussed in
the previous  section, but the minima from the two-center
interference are still very clear. That is because the two-center
interference condition, Eq.~(\ref{sstate}), is given for
field-dressed initial bound states \cite{DressedStates}. In this
case, it is independent on the ellipticity. Similarly, for the
$p$-state contributions, the two-center interference minima appear
in the middle of the each panel in Figs.~\ref {ati_sp}(b) and (d),
because the corresponding electron momentum is perpendicular to the
molecular axis, i.e., $\mathbf{p}\cdot \mathbf{R}=0$. Hence, the
interference condition, Eq.~(\ref{pstate}), always holds.

However, when we consider the full $3\sigma _{g}$ HOMO of the
N$_{2}$ molecule and include $s-p$ mixing, the minima from the
two-center interference can not be observed in the ATI spectra. This
holds for linearly and elliptically polarized fields, as shown in
Figs. \ref{fig1}(a) and (d), and even if only the contributions of
individual orbits are taken, as illustrated in Figs.~\ref{fig1}(c),
(b), (e), and (f). There are two main reasons for it. First, if
$s-p$ mixing is included, the interference condition also depends on
the sum of $\partial _{\tilde{\mathbf{p}}}\psi _{\alpha
}(\tilde{\mathbf{p}})$ in Eq.~(\ref{prefactor2}), where
$\tilde{\mathbf{p}}=\mathbf{p}+\mathbf{A}(t)$. This influence is
expected to be considerable for $\mathrm{N}_2$, as  its $3\sigma
_{g}$ HOMO has a high degree of $s-p$ mixing. Second, the ionization
time $t$ has a positive imaginary part, which can be related to the
width of the barrier through which the electron must initially
tunnel \cite{Becker2002AdvAtMolOptPhys}. Thus, the sum in
Eq.~(\ref{prefactor2}) is usually not purely real or imaginary, and
therefore, the absolute value of Eq. (\ref{prefactor2}) will never
completely vanish, i.e., no clear minima will be observed in the ATI
spectra.

\begin{figure}[tbp]
\begin{center}
\includegraphics[scale=0.3]{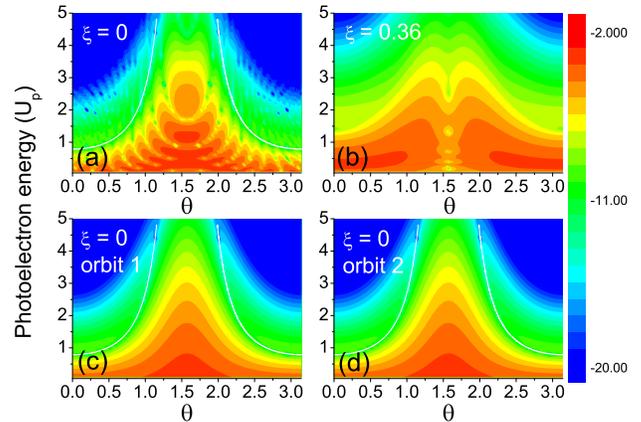}
\end{center}
\caption{(Color online) (a) and (b) Logarithm of ATI spectra of the
N$_{2}$ molecule, coded in false colors in the ($E_{\mathbf{p}}$,
$\protect\theta $) plane. The major axis of the polarization ellipse
is perpendicular to the molecular axis, $\protect\eta=\pi/2$.
Ellipticity $\protect\xi=0$ for (a) and $\protect\xi=0.36$ for (b).
(c) and (d) ATI spectra obtained by employing only individual orbits
in the ($E_{ \mathbf{p}}$, $\protect\theta$) for $\protect\xi=0$.
The white curves indicate the minima caused by the two-center
interference. } \label{ati_sp_90}
\end{figure}

Nevertheless, there is a specific case in which clear spatial
interference fringes are present in the full ATI spectra of N$_2$,
if $s-p$ mixing is included. Fig. \ref{ati_sp_90}(a) shows the
spectra for linear polarization with the alignment angle
$\protect\eta=\pi/2$. Apart from the minima from the temporal
interference, there are two other minima. In order to show them more
clearly, we calculate the ATI transition probability density
$|M_d(\mathbf{p})|^2$ by employing only individual orbits to remove
the pattern associated with the temporal interference.  Figs.
\ref{ati_sp_90}(c) and (d) display the ATI spectra without the
temporal interference, where the clear minima from the spatial
interference are observed. The reason for the appearance of the
spatial interference in this specific case is that for the
perpendicular alignment $\eta=\pi/2$ and for  linear polarization,
the ionization prefactor in Eq. (\ref{prefactor3}) can be simplified
as
\begin{equation}\label{prefactor4}
\begin{split}
d_{ion} = & i \cos(\frac{\mathbf{p}\cdot
\textbf{R}}{2})\sum_{\alpha(s),v }c_{\alpha }b_{v}
\tilde{p}_{x}E_{x}(t) \mathcal{D}_{s}+
\\
& i  \sin(\frac{\mathbf{p}\cdot \textbf{R}}{2})\sum_{\alpha(p),v
}c_{\alpha } b_{v}  \tilde{p}_{x}\tilde{p}_{y}  E_{x}(t)
\mathcal{D}_{p},
\end{split}
\end{equation}
where the momentum $\tilde{p}_{y}=p_{y}$ in the $\hat{\textbf{y}}$
direction is real. Thus, $i\tilde{p}_{x}E_{x}(t)$ can be factorized
in both the first and the second terms in Eq.~(\ref{prefactor4}),
and the remaining terms are time-independent and purely real. Using
Eq. (\ref{prefactor4}), we obtain the minima from the destructive
spatial interference in the ATI spectra, shown by the white curves
in Figs. \ref{ati_sp_90}(a), (c), and (d), which agree well with
what is observed in the spectra. On the other hand, for elliptical
polarization, the imaginary parts in the two terms in Eq.
(\ref{prefactor3}) are different. Hence, they cannot be factorized
and the absolute value of Eq. (\ref{prefactor2}) will never vanish.
This is also the case if the field is linearly polarized, but $\eta
\neq \pi/2$, as seen from Eq.~(\ref{prefactor4}).
Fig.~\ref{ati_sp_90}(b) shows that as the ellipticity increases to
$\xi=0.36$, the minima from both the spatial and temporal
interference in the spectra fade.

\begin{figure}[tbp]
\begin{center}
\includegraphics[scale=0.3]{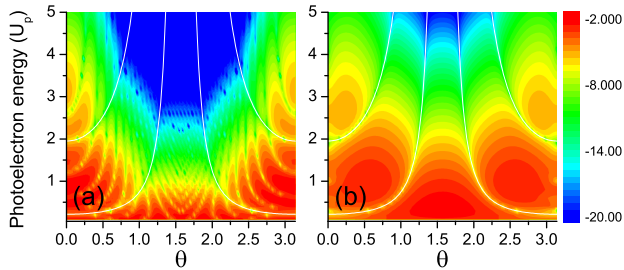}
\end{center}
\caption{(Color online) ATI spectra of the Ar$_{2}$ molecule, with
internuclear distance $R=7.2$ a.u., coded in false colors in the
($E_{\mathbf{p}}$, $\protect\theta$) plane. The major axis of the
polarization ellipse is along the molecular axis, $\eta=0$. The
results in panels (a) and (b) have been computed for ellipticity
$\protect\xi=0$ and $\protect\xi=0.36$, respectively.  The minima
caused by the two-center interference are indicated by white curves.
The remaining field parameters are the same as in Fig.~\ref{fig1}.
}\label{Ar2}
\end{figure}

In contrast, the ATI spectra of the Ar$_{2}$ molecule present clear
minima from the two-center interference, as seen in
Fig.~\ref{Ar2}(a). In addition, Fig. \ref{Ar2}(b) shows that, with
increasing ellipticity, most minima from the temporal interference
are blurred. However, the spatial minima remain unaltered.  The
critical reason for the difference between Ar$_2$ and N$_2$
molecules is that, in the $5\sigma _{u}$ HOMO of Ar$_{2}$, the
contributions of the $p$ states dominate and the contributions of
the  $s$ states are practically vanishing for the internuclear
separation in this work. Hence, the corresponding two-center
interference condition is given by Eq.~(\ref{sstate}), which holds
for \emph{pure} $p$ states and ungerade symmetry. This condition,
shown by the white lines in Fig.~\ref{Ar2}, is independent of the
driving field and is consistent with the interference minima in the
simulation. Furthermore, since Ar$_2$ has a very large equilibrium
internuclear distance, $R=7.2$ a.u., several two-center minima are
observed in the momentum range of interest according to
Eq.~(\ref{sstate}).

\section{Conclusions}

In summary, we theoretically study temporal and spatial interference
in the direct ATI spectra of aligned diatomic molecules in an
elliptically polarized laser field. Temporal interference relates to
ionization events occurring at different times within a field cycle,
and spatial interference is associated with electron ionization at
spatially separated centers in the molecule. Throughout, we have
employed the molecular SFA, saddle-point methods and field-dressed
bound states. We have also assumed that only the HOMO is active.

With increasing ellipticity, there will be a blurring of the
temporal-interference patterns, except when the photoelectron
momenta $\textbf{p}$ are parallel to the major polarization axis.
This blurring is caused by the fact that a nonvanishing driving
field ellipticity affects the electron velocities unequally for the
two ionization events occurring within each field cycle. For linear
polarization, these events are equally probable. In order, however,
to compensate the motion induced by the small component of the
elliptically polarized field, it is necessary that, for one of the
orbits, the electron be released in the continuum with a larger
velocity. This will lead to a suppression of the corresponding
ionization probability, and hence to a disappearance of the temporal
interference patterns. Nevertheless, if the electron is released
along the major polarization axis, the absolute value of the initial
velocity will be the same for the two ionization events. Hence, the
temporal-interference minima in the ATI spectrum will remain clear.
For these specific parameters, we have analyzed in detail how a
non-vanishing ellipticity shifts these minima towards lower
energies, and have traced them to a decrease in the phase difference
between the transition amplitudes related to both orbits. These
results are in agreement with the studies performed in
\cite{Paulus1998PRL}.

Our results also show that, in general, the spatial, two center
interference is much more robust with regard to the electric field
parameters than the temporal interference, as long as the HOMO does
not exhibit a high degree of $s-p$ mixing. As a consequence of $s-p$
mixing, however, or of the temporal dependence of the ionization
prefactor Eq.~(\ref{prefactor2}), the interference patterns become
blurred, even for linear polarization. Only for the very specific
case of linear polarization and a molecule aligned perpendicular to
the field direction will sharp fringes be observed. This is
exemplified by computations in N$_2$. On the other hand, if either
$s$ or $p$ states are dominant in the HOMO, the patterns will be
very clear over a large parameter range and practically independent
of the driving-field ellipticity. This is shown for the ATI spectra
of Ar$_{2}$, which display clear spatial-interference patterns due
to fact that the $p$-state contributions are dominant. Several
minima exist due to the large equilibrium internuclear distance.

Therefore, one can trace all the interference patterns observed, or
the absence thereof, to the dynamics of the electronic wave packet
and the velocity with which it is released in the continuum.  These
patterns can be controlled by an appropriate choice of the the field
and molecular parameters. Concrete examples provided in this work
include the laser-field polarization, the molecular targets, the
angle between the electron momentum and the major polarization axis,
and the molecular orientation with regard to the field. This control
is related to the energy position and to the sharpness of the
patterns. We also identify a parameter region, for which clear
interference fringes can be easily observed. We expect this
information to be useful for future experiments.

\section{Acknowledgement} We are thankful to B. B. Augstein for
his help with GAMESS-UK. This work has been funded by the UK EPSRC
(grant EP/J019240/1), NNSF of China (grant No.~11204356), and the
CAS overseas study and training program.


\begin{thebibliography}{9}

\bibitem{Agostini1979PRL} P. Agostini, F. Fabre, G. Mainfray, G. Petite, and N. K. Rahman, Phys. Rev. Lett. \textbf{42}, 1127 (1979).

\bibitem{Meckel2008Science} M. Meckel \emph{et al.}, Science \textbf{320}, 1478
(2008).

\bibitem{Kang2010PRL} H. Kang  \emph{et al.}, Phys. Rev. Lett. \textbf{104}, 203001
(2010).

\bibitem{Becker2002AdvAtMolOptPhys} W. Becker, F. Grasbon, R. Kopold, D. B. Milo\v{s}evi\'c, G. G. Paulus, and H. Walther, Adv. At. Mol. Opt. Phys. \textbf{48},
35 (2002).

\bibitem{Milo2012PRA} M. Busulad\v{z}i\'{c} and D. B. Milo\v{s}evi\'{c}, Phys. Rev. A \textbf{82}, 015401 (2010).

\bibitem{Bauer2005PRA} D. Bauer, D. B. Milo\v{s}evi\'{c}, and W. Becker, Phys. Rev.
A \textbf{72}, 023415 (2005).

\bibitem{Korneev2012PRL} Ph. A. Korneev \emph{et al}., Phys. Rev. Lett. \textbf{108}, 223601 (2012).

\bibitem{Fano1966PR} H. D. Cohen and U. Fano, Phys. Rev. \textbf{150}, 30
(1966).

\bibitem{Walter1999JPB} M. Walter and J. Briggs, J. Phys. B \textbf{32}, 2487
(1999).

\bibitem{Lei2002PRA} M. Lein, J. P. Marangos, and P. L. Knight, Phys. Rev. A \textbf{66}, 051404(R) (2002).

\bibitem{selsto2005PRL} S. Selst{\o}, M. F{\o}rre, J. P. Hansen, and L. B.
Madsen,  Phys. Rev. Lett. \textbf{95}, 093002 (2005).

\bibitem{Baltenkov2012JPB} A. S. Baltenkov, U. Becker, S. T. Manson, and A. Z. Msezane, J. Phys. B \textbf{45}, 035202 (2012).

\bibitem{bohm2000PRL} J. Muth-B\"{o}hm, A. Becker, and F. H. M.
Faisal,  Phys. Rev. Lett. \textbf{85}, 2280 (2000).


\bibitem{Busuladzic2008PRA} M. Busulad\v{z}i\'c. A. Gazibegovi\'c-Busulad\v{z}i\'c, D. B.
Milo\v{s}evi\'c, and W. Becker, Phys. Rev. A \textbf{78}, 033412
(2008).

\bibitem{Carla2011PRA} T. Shaaran, B. B. Augstein, and C. Figueira de Morisson
Faria, Phys. Rev. A \textbf{84}, 013429 (2011).

\bibitem{Li2012PRA} W. Li and J. Liu, Phys. Rev. A \textbf{86}, 033414
(2012).

\bibitem{ottawa06} N. Dudovich, J. Levesque, O. Smirnova, D. Zeidler, D. Comtois, M. Yu. Ivanov, D. M. Villeneuve, and P.
B. Corkum, Phys. Rev. Lett. \textbf{97}, 253903 (2006).

\bibitem{Lai2013PRL} X. Y. Lai \emph{et al}., Phys. Rev. Lett. \textbf{110}, 043002 (2013).

\bibitem{Eckle2008NatureP} P. Eckle \emph{et al.}, Nature Phys. \textbf{4}, 565 (2008).

\bibitem{Goreslavski2004PRL} S. P. Goreslavski, G. G. Paulus, S. V. Popruzhenko, and
N. I. Shvetsov-Shilovski, Phys. Rev. Lett. \textbf{93}, 233002
(2004).

\bibitem{Staudte2009PRL} A. Staudte \emph{et al}., Phys. Rev. Lett. \textbf{102}, 033004
(2009).

\bibitem{Busuladzic2009PRA} M. Busulad\v{z}i\'c. A. Gazibegovi\'c-Busulad\v{z}i\'c, and D. B.
Milo\v{s}evi\'c, Phys. Rev. A \textbf{80}, 013420 (2009).

\bibitem{Paulus1998PRL} G. G. Paulus, F. Zacher, H. Walther, A. Lohr, W. Becker and M. Kleber, Phys. Rev. Lett. \textbf{80}, 484 (1998).

\bibitem{Carla2002PRA} C. Figueira de Morisson Faria, H. Schomerus, and W.
Becker, Phys. Rev. A \textbf{66}, 043413 (2002).

\bibitem{KFR} L. V. Keldysh, Zh.Eksp. Teor. Fiz. \textbf{47}, 1945 (1964)
[Sov. Phys. JETP \textbf{20}, 1307 (1965)]; F. H. M. Faisal, J.
Phys. B \textbf{6}, L89 (1973); H. R. Reiss, Phys. Rev. A
\textbf{22}, 1786 (1980).

\bibitem{Carla2010PRA} C. Figueira de Morisson Faria and B. B. Augstein, Phys. Rev. A \textbf{81}, 043409 (2010).

\bibitem{Milo2006PRA} D. B. Milo\v{s}evi\'{c}, Phys. Rev. A \textbf{74}, 063404 (2006).

\bibitem{Madsen2004JPB} T. K. Kjeldsen and L. B. Madsen, J. Phys. B \textbf{37}, 2033
(2004).

\bibitem{Chu2005PRA} V. I. Usachenko and Shih-I. Chu, Phys. Rev. A \textbf{71}, 063410 (2005).
\bibitem{Lein_Review} M. Lein, J. Phys. B \textbf{40}, R135 (2007).
\bibitem{Carla_Review} B. B. Augstein and C. Figueira de Morisson Faria, Modern Phys. Lett. B \textbf{26}, 1130002 (2011).
\bibitem{Madsen2006} C. B. Madsen and L. B. Madsen, Phys. Rev. A \textbf{74}%
, 023403 (2006)

\bibitem{Wahl1964JCP} A. C. Wahl, J. Chem. Phys. \textbf{41}, 2600 (1964).

\bibitem{GamessUK} GAMESS-UK is a package of \emph{ab initio} programs. See
http://www.cfs.dl.ac.uk/gamess-uk/index.shtml, M. F. Guest, I. J.
Bush, H. J. J. van Dam, P. Sherwood, J. M. H. Thomas, J. H. van
Lenthe, R. W. A. Havenith, J. Kendrick, Mol. Phys. \textbf{103}, 719
(2005).
\bibitem{Bleistein} N. Bleistein and R. A Handelsman, \textit{``Asymptotic Expansions of Integrals"} (Dover, New York, 1975)
\bibitem{KopoldPhD} R. Kopold, PhD thesis, Technical University Munich,
Chapt. 5 (2001).

\bibitem{SF2010} T. Shaaran and C. Figueira de Morisson Faria, J. Mod. Opt.
\textbf{57}, 984 (2010); T. Shaaran, M. T. Nygren, and C. Figueira
de Morisson Faria, Phys. Rev. A \textbf{81}, 063413 (2010).

\bibitem{SFS2012} T. Shaaran, C. Figueira de Morisson Faria, and H.
Schomerus, Phys. Rev. A \textbf{85}, 023423 (2012).

\bibitem{DressedStates} D. B. Milo\v{s}evi\'{c}, Phys. Rev. A \textbf{74},
063404 (2006); W. Becker, J. Chen, S. G. Chen, and D. B.
Milo\v{s}evi\'{c}, Phys. Rev. A \textbf{76}, 033403 (2007).

\bibitem{Delone1991JOSA} N. B. Delone and V. P. Krainov, J. Opt. Soc. Am. B
\textbf{8}, 1207 (1991).

\bibitem{Goreslavski1996ZETF} S. P. Goreslavski and S. V. Popruzhenko, Zh.
Eksp. Teor, Fiz. \textbf{110}, 1200 (1996) [Sov. Phys. JETP
\textbf{83}, 661 (1996)].

\end{thebibliography}
\end{document}